\documentstyle[epsfig,12pt]{article}
\def\beq{\begin{equation}}
\def\eeq{\end{equation}}
\def\ber{\begin{eqnarray}}
\def\eer{\end{eqnarray}}
\def\ben{\begin{enumerate}}
\def\een{\end{enumerate}}
\def\ni{\noindent}

\def\beb{\begin{thebibliography}{99}}
\def\eeb{\end{thebibliography}}
\def\bi{\bibitem{}}

\newcommand{\qbar}{\overline{q}}
\begin{document}
\ni {\bf {\large $qq^{\prime} \rightarrow qq^{\prime}$} : A SECOND LOOK AT THE IR-DIVERGENCES} \\

{\hspace{2.00cm} \sl Sushan Konar$^{\dagger,\ddagger}$\footnote{sushan@rri.ernet.in, sushan@physics.iisc.ernet.in}, 
Avijit Kumar Ganguly$^{\ast}$\footnote{avijit@cts.iisc.ernet.in}}\\
\centerline{\small $^{\dagger}$Astrophysics Group, Raman Research Institute, Bangalore}\\
\centerline{\small $^{\ddagger}$JAP, Department of Physics, Indian Institute of Science, Bangalore}\\ 
\centerline{\small $^{\ast}$Centre for Theoretical Studies, Indian Institute of Science, Bangalore}\\ 
\baselineskip=11pt

\ni Interest in the transport properties of the quark-gluon plasma has arisen mostly keeping in mind two 
widely differing kinds of systems - that of the high temperature phase created in the Relativistic Heavy Ion
Colliders and the effectively zero temperature, Fermi-degenerate phase believed to exist in the cores 
of the Neutron Stars or in the more exotic Strange Stars. In the first case, the transport coefficients (thermal
and electrical conductivities and shear viscosity) determine the nature of the equlibration. Whereas in the 
case of the astrophysical systems the cooling history, the rotational behaviour and the evolution of the magnetic 
field depend on such coefficients, and may even provide an answer to the question whether some of the pulsars 
are indeed Strange Stars [1].\\

\ni To this end, accurate calculations of the {\em elastic scattering cross-sections} are required for 
$qq^{\prime} \rightarrow qq^{\prime}$ and $ q\qbar^{\prime} \rightarrow q\qbar^{\prime}$ processes,
which could then be used to solve the Boltzman transport equation to yield the required transport
coefficients. Such calculations have already been performed at the tree-diagram level where exchange 
of massless gluons leads to divergent cross sections [2]. This is analogous to Coulomb scattering 
where transverse photons are exchanged in an electron plasma. For small momentum transfer in $qq^{\prime}$ 
scattering, the cross-section is power-divergent just like Rutherford cross-section 
($\frac{d\sigma}{d\Omega} \sim \frac{g^4}{q^4}$). Taking the actual amount of momentum transfer into account 
the cross-section is found to be log-divergent, which is still short of curing the malady.\\

\ni Debye screening, removes the IR-divergences for the electric or the longitudinal modes by providing an
effective length scale to the system. And the current-current interaction term provides 
a dynamic screening akin to Landau damping for the transverse modes. An effective way to implement this
screening is to intoduce a Debye mass as an infrared regulator into the gluon propagator for the thermal
gluons. This Debye mass cuts off the magnetic divergence too [3] and the dynamic screening is taken care of.\\

\ni But, to our dismay, we find that at the two-gluon exchange level such a technique fail to remove the divergences. 
This constitutes the main result of our work that we would like to report here. In all the previous work, though 
a thermal mass has been introduced for the gluon propagator to take care of the in-medium effects, but the external 
fermions (the quarks) are still taken to be massless, in the high-temperature as well as in the effective 
zero-temperature phase.\\

\ni To demonstrate the divergent behaviour in the fermion-fermion scattering cross section with two gluon exchange, 
we take the external quark lines to be massless and use the following forms for the propagators and the vertex 
structures [4], \\
i) $g f^{abc}[(p-q)_{\nu} g_{\lambda \mu} + (q-r)_{\lambda} g_{\mu \nu} + (r-p)_{\mu} g_{\nu \lambda}]$
- three gluon vertex,\\
ii) $i g \gamma^{\mu} T_{ij}^a$ - the one gluon - two quark vertex, \\
iii) $\frac{1}{\not{p + i\epsilon}}$ - the fermion propagator,\\
iv) $D^{ab}_{\mu \nu}(k) = g_{\mu \nu} \delta^{ab} [ \frac{i}{k^2 - m_g^2 + i \epsilon}
+ 2 \pi \delta(k^2 - m_g^2) n(k)]$ - the thermal gluon propagator, 
where, $n(k)$ is the Bose-Einstien distribution function and the thermal gluon mass 
is given by $m_g^2 = \frac{1}{6} C_A g^2 T^2 + \frac{1}{12} C_F g^2 (T^2 + 3 \frac{\mu^2}{\pi^2})$, $C_F, C_A$ being 
Casimir invariants of the symmetry group in the fundamental representation and the adjoint representation. 
$\mu$ is the chemical potential of the quarks and we work in the regime where $T << \mu$.\\

\begin{figure}
\epsfig{file=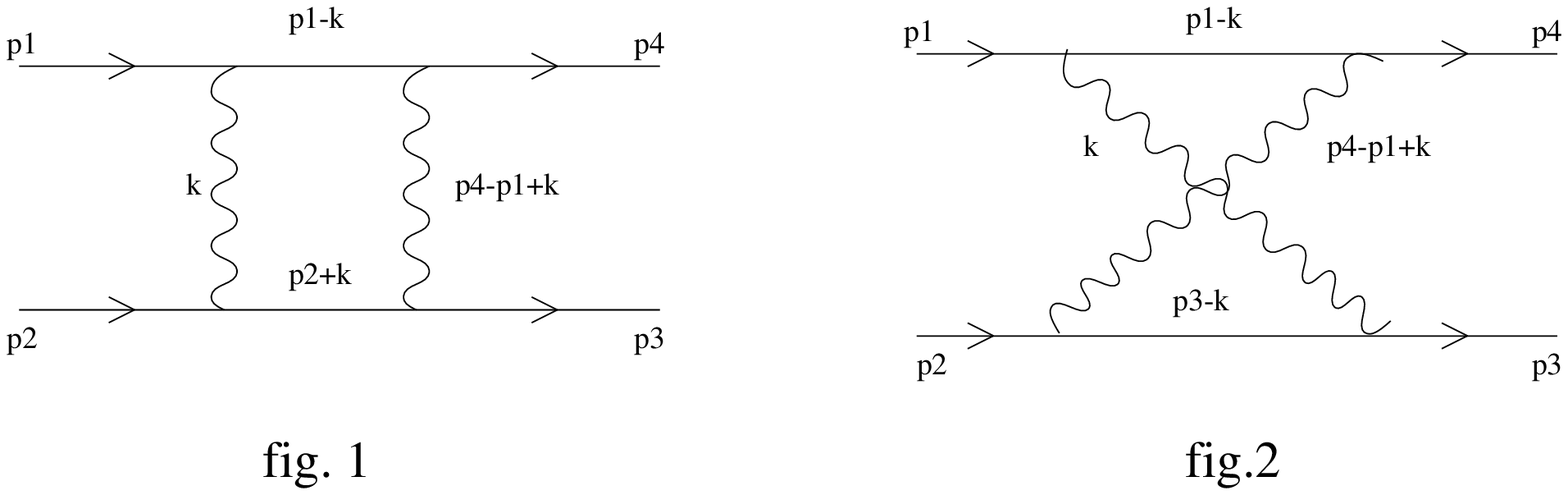,width=250pt}
\end{figure}

\ni To show the IR-divergence of the scattering cross-section in the above-mentioned situation it is enough to 
consider just the t-channel process. For such a process the diagrams could be classified into three categories, 
(i) the exchange type similar to those arising in QED, (ii) diagrams where self-energy corrections are included on 
the external legs, (iii) diagrams with self-energy corrections on the exchanged gluons and (iv) the vertex correction 
types. It is only the diagrams of type (i) that suffer from the infrared divergences. Therefore, here we concentrate 
on the diagrams of type (i) particularly those shown in fig.1 and fig.2, since the other diagrams of this variety 
are suppressed by factors of $s$ ( the square of the sum of initial momenta). \\

\ni At the two-gluon exchange level, the leading contribution to the amplitude, in the limit $t \rightarrow 0$ 
($t$ is the square of the momentum transfer) comes from the thermal part of the two gluon exchange lines, particularly 
from the delta functions. So retaining the leading term in $\frac {1}{t}$, the matrix element for these diagrams is 
given by,
\ber
<{|M_{qq^{\prime}}|}^2> &=& \frac{1}{9} \Sigma_c (2 \pi g^2)^2
\frac{1}{4} \Sigma_{\rm spin} {|fig.2a + fig.2c|}^2 \nonumber \\
&=& \frac{1}{9} \Sigma_c (2 \pi g^2)^2 \frac{1}{4} \Sigma_{\rm spin}
|(\Gamma^{\mu}_4 \gamma^{\alpha} \gamma^{\nu} \Sigma_{13} \gamma_{\mu} \gamma^{\beta} \Gamma_{2 \nu})
\frac{T}{\sqrt{s} (2 \pi)^3 (2 m_g^2 + s) t}  \nonumber \\
& & (p_{1 \alpha} {(p_1 - p_2)}_{\beta} - p_{4 \alpha} {(p_4 - p_3)}_{\beta}) \nonumber \\
& & + (\Gamma^{\mu}_4 \gamma^{\alpha} \gamma^{\nu} \Sigma_{13} \gamma_{\nu} \gamma^{\beta} \Gamma_{2 \mu})
\frac{T}{\sqrt{s} (2 \pi)^3 (2 m_g^2 + u) t} \nonumber \\
& & (p_{1 \alpha} {(p_1 + p_3)}_{\beta} - p_{4 \alpha} {(p_4 + p_2)}_{\beta})|^2
\eer
where, $p_1, p_2$ are the incoming momenta, $p_3, p_4$ are the outgoing momenta with,
\ber
s = (p_1 + p_2)^2, u = (p_1 - p_3)^2, t = (p_1 - p_4)^2, \\
\Gamma^{\mu}_4 = \overline{U}(p_4) \gamma^{\mu}, \Sigma_{13} = U(p_1) \overline{U}(p_3)
\eer
and $T$ is the temperature. Where we have, $s + u + t = 0$, as $s, u, t$ are the Mandelstam variables.
The matrix element is obtained by summing over the final states and averaging over the initial states.
The finite temperature phase space measures are given by,
\ber
d\phi_4 &=& \frac{d^4 p_4}{(2 \pi)^4} \delta(p_4^2 - m_0^2) [ \Theta(p_{40}) - N (|p_{40}|)] \\
d\phi_3 &=& \frac{d^4 p_3}{(2 \pi)^4} \delta(p_3^2 - m_0^2)
[\Theta(p_{30}) - N (|p_{30}|)]
\eer 
Here $m_0 = 0$ as the external quarks are massless and the factor $[\Theta(p_{40}) - N (|p_{40}|)]$ takes care of 
Pauli blocking [5]. \\

\ni Then the total scattering cross-section is given by
\ber
\sigma(i \rightarrow f) &=& \int \frac{1}{p_1.p_2} <{|M|}^2> \delta(p_1 + p_2 - p_3 - p_4) (2 \pi)^4 \nonumber \\
& & \frac{d^4 p_4}{(2 \pi)^4} \delta(p_4^2) [ \Theta(p_{40}) - N (|p_{40}|)]
\frac{d^4 p_3}{(2 \pi)^4} \delta(p_3^2) [ \Theta(p_{30}) - N (|p_{30}|)] \nonumber \\
&=& \frac{1}{s} \int \frac{d\Omega_{p_4}}{(2 \pi)^4} <{|M|}^2> [1 - N(\frac{\sqrt{s}}{2})]^2 \nonumber \\
&\sim& \frac{1}{s^2} <{|M|}^2> [1 - N(\frac{\sqrt{s}}{2})]^2
\eer
as, $d\Omega = \frac{d\phi d(-t)}{s/2}$.\\
Therefore, 
\beq
\frac {d \Sigma}{dt} \sim {\frac {1}{4 {\pi}^3}}{\frac {1}{s^2}} < {|M|}^2 > \left[ 1 - N (\sqrt s ) \right]^2
\eeq
\\

\ni Since, to leading order, $<|M|^2> \sim O(\frac{1}{t}) + ..$, which has a {\em log divergence} in small momentum 
transfer we conclude that {\em infra-red divergence is NOT cured in two-gluon exchange level by introducing an effective
mass to the thermal gluons only.} As $\sigma$ diverges for small $t$, we once again have the situation of vanishing 
transport coefficients! One is saved from such absurd results by considering non-zero mass for the external quarks. 
In a chiral symmetry broken phase one could introduce the mass following the standard procedure and that cures the 
problem. But in a chirally symmetric phase one needs to consider the chirally invariant mass of the fermions [6]. 
A detailed report of this work will be presented elsewhere [7].\\

\ni We'd like to thank Prof. McLerran for helpful discussions, Dr. Indumathi for kindly going through our calculations,
and Dr. Prakash Mathews for a critical reading of the manuscript.

\beb
\baselineskip=10pt
\bi S. Konar, in preparation
\bi M.~H. Thoma, Phys.Rev., {\underline {D49}}, 451,1994
\bi G. Bayam, H. Monien, C.~J. Pethik and D.~G. Ravenhall, Phys.Rev.Lett., {\underline {64}}, 1867, 1990
\bi M. Labellac, {\em Thermal Field Theory}, Cambridge University Press, 1996
\bi H.~A. Weldon, Phys.Rev., {\underline {D26}}, 2789 ,1982.
\bi H.~A. Weldon, Phys.Rev., {\underline {D49}}, 1579 ,1994.
\bi A.~K. Ganguly and S. Konar, in preparation
\eeb

\end{document}